\documentstyle[preprint,aps]{revtex}
\begin{document}
\draft
\title{Gauged motion in general relativity and in Kaluza-Klein theories}
\author{Mohammad~Nouri-Zonoz$^{a,b,c}\thanks{Electronic address:~nouri@theory.ipm.ac.ir}$ and
Ali~Reza~Tavanfar$^{a,b}\thanks{Electronic
address:~art@ipm.ir}$ }
\address{$^{a} $ Department of Physics, University of Tehran, End of North Karegar St.,
14352 Tehran, Iran.\\
$^{b}$ Institute for studies in theoretical physics and
Mathematics, P O Box 19395-5531 Tehran, Iran.\\
$^{c}$ Institute of Astronomy, Madingley Road, Cambridge, CB3 0HA, UK.}
\maketitle
\begin{abstract}
In a recent paper [1] a new generalization of the Killing motion, the
{\it gauged motion}, has been introduced for stationary spacetimes
where it was shown that the physical symmetries of such spacetimes
are well described through this new symmetry.
In this article after a more detailed study in the stationary case we present
the definition of gauged motion for general spacetimes. The definition is based on the
gauged Lie derivative induced by a threading family of observers and the relevant
reparametrization invariance. We also extend the gauged motion to the case of
Kaluza-Klein theories.
\end{abstract}
\pacs{PACS No.}
$$ \wp\wp\wp\wp\wp\wp\wp\wp\wp\wp\wp\wp\wp\wp\wp\wp\wp\wp\wp\wp\wp\wp\wp\wp\wp\wp\wp\wp\wp\wp\wp\wp\wp $$
\section*{Contents}
$$ $$
$1.\;\;\;\;${\bf Introduction} \\
$2.\;\;\;\;${\bf Orbit manifolds and the parametric structure}\\
$3.\;\;\;\;${\bf Summary on the $1+3$ decomposition in general
relativity}\\
$4.\;\;\;\;${\bf Gauged Lie derivative}\\
$5.\;\;\;\;${\bf Gauged motion in the stationary case}\\
$6.\;\;\;\;${\bf Curvature invariants of stationary spacetimes and the gauged
motion}\\
$7.\;\;\;\;${\bf NUT-type spaces and their symmetries}\\
$8.\;\;\;\;${\bf Some general properties}\\
$9.\;\;\;\;${\bf Gauged motion in general spacetimes}\\
$10.\;\;\;${\bf Kaluza-Klein theories and the extended gauged motion}\\
$11.\;\;\;${\bf Comparison with the case in QFT}\\
$12.\;\;\;${\bf Conclusions}\\
$A.\;\;\;\;${\bf Gravitoelectromagnetic fields, the spatial force
and geometry of splitting}\\
$$ \wp\wp\wp\wp\wp\wp\wp\wp\wp\wp\wp\wp\wp\wp\wp\wp\wp\wp\wp\wp\wp\wp\wp\wp\wp\wp\wp\wp\wp\wp\wp\wp\wp $$
\section{Introduction}
The well known definition of a spacetime symmetry is based on the
concept of \textit{isometry} and formulated as the Killing equation
(or Killing motion)
$${\pounds_\xi} g_{a b} = 0. \eqno(1.1)$$
However applying the concepts of the {\it threading} approach (section {\bf III})
to the spacetime decomposition one can show that the physical symmetries of a
spacetime corresponding to a given timelike family of observers might be different
from its apparent mathematical symmetries exhibited by the metric of
that spacetime [1]. This difference is due to
the fact that these spacetime symmetries are the motions under which
the observer clock rates, the spatial metric $h_{ab}$ and the
tensor fields appearing in the spatial force, specially the gravitoelectromagnetic
 fields ${\cal E}_{a}$ and ${\cal B}_{ab}$, are invariant
and therefore there exists a non sensitivity to the gravitomagnetic
potential $A_{a}$ up to a suitable \textit{gauge} transformation [1,2].
A known example is the NUT spacetime [3], although the metric itself dose not have
spherical symmetry but
the spacetime really does, in fact one can show that all the curvature invariants of the
spacetime are spherically symmetric [4,5]. The same property have
been shown for the cylindrical (planar) symmetry through the so called
cylindrical (planar) analogue of NUT space [6,1].
The outline of the paper is as follows. After a brief
mathematical preliminary on orbit manifolds and the relevant splitting structure
in the next section, we give a summary on the $1+3$
formalism of the spacetime decomposition in section {\bf III}. Concepts introduced
in this section will be used extensively throughout the paper.
In section {\bf IV}, on the basis of a reparametrization invariance of the threading decomposition,
 we define a generalization of Lie derivative called the \textit {gauged Lie derivative}.
 In section {\bf V} we introduce the idea of \textit {gauged motion} or \textit{gauged isometry}
  in the stationary spacetimes and in section {\bf VI} we show that in these
 spacetimes the gauged isometries are in accordance with the symmetries of the curvature invariants. Section {\bf VII} is devoted to our primary motivation of introducing the idea of gauged motion i.e providing a clear and covariant manifestation of the hidden symmetries of NUT space which had been a matter of discussion for sometime. Some of the general properties of
\textit{gauged Killing vectors}, (GKV), in stationary case are discussed in section {\bf VIII}.
 In section {\bf IX} the gauged motion is defined for general
spacetimes according to the ideas of sections {\bf II}, {\bf III} and {\bf IV}.
 Then considering Kaluza-Klein theories and symmetries of the corresponding 4-D
spacetimes, an extended gauged motion is presented is section {\bf X} for such theories.
In section {\bf XI} we make a brief comparison between spacetime symmetries through gauged motion
and spontaneous symmetry breaking in
quantum field theory. Conclusions and further applications are summarized in section {\bf XII}.
There is also an appendix on the derivation of the spatial force
and the geometrical interpretations of the gravitoelectromagnetic fields.
\section{Orbit manifolds and the parametric structure}
let $\Re=({\cal R},+)$ the additive Lie group of real numbers,
act smoothly on a (pseudo)-Riemannian manifold $(M,g)$ on
the left. That is a smooth map
$$\varrho: {\cal R} \times M \rightarrow M$$
satisfying
$$\varrho(\tau_{2} + \tau_{1},q)= \varrho(\tau_{2},\varrho(\tau_{1},q))\;\;
;\;\;\forall(\tau_{1},\tau_{2},q)\in {\cal R}^{2} \times M$$
$$\varrho(0,q)=q\;\;;\;\;\forall q \in M$$
defines a global flow on $M$ or in other words \textit{threads}
it. Such a structure is called a (smooth) $\Re$-manifold.\footnote{If $M$ is taken to be a state
space, this is the definition of
a smooth dynamical system of $M$.} Now the set of maps
$$\varrho_{\tau}:\;M \rightarrow M\;\;;\;\;\forall \tau \in {\cal R}$$
$$\varrho_{\tau}(q)=\varrho(\tau,q)$$
with the corresponding composition as the group operation forms a 1-parameter group
of \textit{diffeomorphisms} of $M$ and the orbits
$${\Re}.q= \{\varrho(\tau,q)\;\;;\;\;\forall \tau \in {\cal R}\}
\;\;;\;\;\forall q \in M$$
form a congruence of one dimensional immersed smooth submanifolds of $M$ on $M$.
Moreover the corresponding stabilizer of each point $q\in M$
$$\Re_{q}=\{\tau\in {\cal R}\;\;|\;\;\varrho(\tau,q) = q\;\}$$ is a closed subgroup of $\Re$ and the left coset space $\frac{\Re}{\Re_{q}}$ can be identified with $\Re.{q}$ [7].
Therefore due to the fact that the only closed subgroups of $\Re$
are
$$({0},+)\;\;,\;\;\Re\;\;,\;\;(\{n\tau;\forall n\in{\cal Z}\},+);\;\tau\in {\cal R}$$
each orbit could be diffeomorphic to ${\cal R}$, $S^{1}$ or a point but here all
the orbits are assumed to be either ${\cal R}$ or $S^{1}$. Denoting such an action by the
triplet $(\varrho,\Re,M)$ and considering the orbit space
$$\bar{M} = \frac{M}{\Re} = \{\Re.q\;\;;\;\; \forall q \in M\}$$
together with the quotient topology, i.e. the largest topology which makes the map
$$\Pi: M \rightarrow \bar{M}\;\;;\;\; {\Pi(q)}={\Re}.q $$
everywhere continuous, one may now raise the question whether $\bar{M}$ could be
given the structure of a smooth manifold. The answer is in the affirmative and
the condition needs to be fulfilled by the smooth action on $M$ are explained in the following theorem.\\
\textbf{\textit{Theorem}}:\\
Let $G$ be a Lie group acting smoothly on a smooth manifold $M$ (on the left)\footnote{Here a manifold is defined as a Hausdorff, second countable topological space which is locally homeomorphic
to a finite dimensional Euclidean space.}. The topological orbit space $\bar{M}=\frac{M}{G}$ has the structure of a smooth
manifold with $\Pi:
M \rightarrow \bar{M}$ a submersion if and only if
$$\sim=\{(q,q^{'})\in M\times M\;\;|\;\;\Pi(q)=\Pi(q^{'})\}$$
is a closed smooth submanifold of $M \times M$ with respect to its product topology.
The manifold structure on $\bar{M}$ satisfying these requirements is
then unique, moreover if $M$ and $\Pi$ are analytic, so is $\bar{M}$ [8].\\
In the case of a compact Lie group $G$, however a sufficient condition is that the action be free, i.e the stablizer of each point of $M$ is the identity subgroup of $G$ [9].\\
In what follows we assume that, apart from the analyticity, the above theorem is satisfied.
All the above arguments
could be simply encoded in the fact that either of the triplets $(M,\Re,\bar{M})$  or
$(M,U(1),\bar{M})$ is a \textit{smooth principal bundle}.\\
There is a unique non-null vector field on $M$ which generates $(\varrho,\Re,M)$.
It is defined by\footnote{Our notation is such that given a vector field $X$ and a point
$q\in M$, ${}_{q}X \equiv X(q)\;$.}
$$\zeta: M \rightarrow TM $$
$$_{q}\zeta({\rm f})=\frac{d}{d\tau}_{|\tau=0}{\rm f}[\varrho_{\tau}(q)] $$
where ${\rm f}$ is any real-valued function defined on a neighborhood of the point.
Defining the curves
$$_{q}\Gamma: I=(-1,1) \rightarrow M\;\;;\;\;\forall q \in M$$
$$_{q}\Gamma[{\varphi}(\tau)] = \varrho(\tau,q)$$
with $\varphi\;: {\cal R} \rightarrow I$ a homeomorphism,
$_{q}\zeta$ is  the tangent vector to $_{q}\Gamma$ at the point
$q$.\\
There is a unique \textit{orthogonal splitting structure} on $(M,g)$, corresponding
to $(\varrho,\Re,M)$, constructed by using the projection $(P)$ and coprojection $(P^{'})$
tensor fields, defined as
$$P^{'}(\omega,X) = - \zeta \otimes g\;(\omega,\zeta,X)\;\;;\;\;\forall (X,\omega) \in
TM \times T^{*}M \eqno(2.1.a)$$
$$P = \delta - P^{'} \eqno(2.1.b)$$
where
$$\delta(\omega,X) = \omega(X)\;\;;\;\;\forall (X,\omega)\in TM \times T^{*}M.$$
They satisfy the relations
$$P^{2}=P\;\;;\;\;P^{'2}=P^{'}\;\;;\;\;PP^{'}=P{'}P=0\;\;;\;\;P + P^{'}=\delta $$
and split the tangent bundle over $M$ to a direct sum
$$ TM =\;{{}^{\perp}TM}\;{\oplus}\;{{}^{\parallel}TM}  \eqno(2.2.a)$$
in which
$${}^{\perp}TM =\;{\cup_{q\in M}}\;{}^{\perp}T_{q}M   $$
$${}^{\perp}T_{q}M=\;\{{}_{q}X\;\; ;\;\;{}_{q}X.{}_{q}\zeta=0\}{}\;\;;\;\;\forall q \in M. $$
Similarly
$$T^{*}M =\;{{}^{\perp}T^{*}M}\;{\oplus}\;{{}^{\parallel}T^{*}M}. \eqno(2.2.b)$$
The splitting relations $(2.2.a)$ and $(2.2.b)$ are based on the following unique decompositions
$$ X\;=\;{}^{\perp}X\;+\;{}^{\parallel}X\;\;;\;\;\forall X \in TM$$
$$\omega\;=\;{}^{\perp}\omega\;+\;{}^{\parallel}\omega{}
\;\;;\;\;\forall \omega \in T^{*}M$$
where $$\omega({}^{\perp}X)\;=\;{}^{\perp}\omega(X)\;=\;P(\omega,X)\;\;;\;\;
\forall (X,\omega) \in TM \times T^{*}M. \eqno(2.3)$$
In this respect the projected tensor fields
$${}^{\perp}T \in\;{}^{\perp}TM^{r}_{s}\;\;;\;\;\forall T \in TM^{r}_{s}$$
are defined by
$$ ^{\perp}T({}^{\perp}\omega_{1},...,{}^{\perp}\omega_{r},...,
{}^{\perp}X_{1},...,{}^{\perp}X_{s})\;=\;
T({}^{\perp}\omega_{1},...,{}^{\perp}{\omega}_{r},...,
{}^{\perp}X_{1},...,{}^{\perp}X_{s})$$
$$;\;\;\forall(\omega_{1},...,\omega_{r},X_{1},...X_{s})
\in T^{*}M^{r}\times TM^{s}.    \eqno(2.4) $$
Therefore the metric tensor of the orthogonal tangent bundle, ${}^{\perp}TM$, is
$$\bar{g} ={}^{\perp}g. $$
If $\bigtriangledown$ is an affine connection on $TM$ then
$${}^{\perp}{\bigtriangledown}_{X}X^{'}\;=\;{}^{\perp}
({\bigtriangledown}_{X}X^{'})\;\;;\;\;\forall (X,X^{'})
\in {}^{\perp}TM^{2}     \eqno(2.5)$$
is an affine connection on ${}^{\perp}TM$ [10]. Moreover if $\bigtriangledown$ is
 torsion-free
and compatible with $g$, ${}^{\perp}\bigtriangledown$ is torsion-free and compatible with ${}^{\perp}g$ where the two torsions are related to each other by [11]
$${}^{\perp}{\cal T}(X,X^{'}) = {}^{\perp}({\cal T}(X,X^{'}))\;\;;\;\;
\forall (X,X^{'})\in {}^{\perp}TM^{2}.     $$
The so called \textit{Zelmanov} curvature tensor field on ${}^{\perp}TM$ is defined
 through [12,11]
$$ \bar{Z}(X^{'},X^{''})X=
{}^{\perp}{\bigtriangledown}_{X^{'}}{}^{\perp}{\bigtriangledown}_{X^{''}}X-
{}^{\perp}{\bigtriangledown}_{X^{''}}{}^{\perp}{\bigtriangledown}_{X^{'}}X-
^{\perp}(\pounds_{[X^{'},X^{''}]}X)$$$$;\;\;\forall(X,X^{'},X^{''})
\in ^{\perp}TM^{3}. \eqno(2.6)$$
With respect to a given basis one has\footnote{In this paper the Latin
indices run from 0 to d-1 while the Greek ones run from 1 to d-1 for a d-manifold.}
$$P^{a}_{b}=\delta^{a}_{b}-\frac{\zeta^{a}\zeta_{b}}{|\zeta|^{2}} \eqno(2.7.a)$$
$$^{\perp}T^{a_{1}...a_{r}}_{b_{1}...b_{s}}=
P^{a_{1}}_{c_{1}}...P^{a_{r}}_{c_{r}}P^{d_{1}}_{b_{1}}...P^{d_{s}}_{b_{s}}
T^{c_{1}...c_{r}}_{d_{1}...d_{s}}$$
$$\bar{g}_{ab}=g_{ab}-\frac{\zeta_{a}\zeta_{b}}{|\zeta|^{2}}. \eqno(2.7.b)$$
Applying the above definitions to the 4-D holonomic
basis $\partial_{a}$ and its dual ${\rm d}x^{a}$ one gets the 3-D, generally
non-holonomic bases for the orthogonal tangent and cotangent spaces
$$ ^{\perp}{\partial}_{a}= P^{b}_{a} \partial_{b} \eqno(2.8.a)$$
$${}^{\perp}{\rm d}^{a}=P^{a}_{b} {\rm d}x^{b} \eqno(2.8.b)$$
Accordingly the absolute derivative corresponding to
${}^{\perp}{\bigtriangledown}$ is
$${}^{\perp}{\cal D} \doteq {}^{\perp}{\rm d}^{a}\;{}^{\perp}{\bigtriangledown}_{a}. \eqno(2.8.c)$$
\textbf{\textit{Definition}}:\\
A \textit{parametric orbit manifold} $\bar{\cal{M}}$ corresponding to
an $\Re-$manifold $(M,g)$ is the orbit manifold $\bar{M}=\frac{M}{\Re}$
endowed with the tensor fields and
the connection defined on the orthognal tangent and cotangent bundles,
$({}^{\perp}TM,{}^{\perp}T^{*}M)$. [10,12,13]\\
\section{Summary on the $1+3$ decomposition in general relativity}
Due to the coupling of space, time and matter in general relativity
the true reconstruction of the 3-D space in a general spacetime has
been a matter of investigation. Any satisfactory solution to this
problem should respect the following two requirements.\\
$I.)\;$From the geometrical standpoint the definition of space has to be \textit{intrinsic}
 i.e coordinate independent.\\
$II.)\;$From the physical standpoint since there is no notion of absolute time or
absolute space
 in the theory of relativity, the definition has to be somehow \textit{observer-dependent}.\\ To satisfy both of the above requirements one could start from a
 threading family of observer worldlines in a given spacetime.
 Then a standard experiment in general relativity (local light synchronization), based on sending
  and
 receiving light signals, defines the
 \textit{spatial distance} between any two nearby observers of this family [14].
 The \textit{spatial metric tensor} defined in this way
 coincides (up to a minus sign) with the metric tensor ${}^{\perp}g$ of the orthogonal tangent bundle
 induced by the observer worldlines. According to this fact and some other similar arguments,
 the corresponding parametric orbit manifold
 is taken to be the 3-D space \textit{realized} by these observers.
 This method is called the $1+3$ spacetime splitting or the threading
 decomposition [10,15].\footnote{For various applications of this formalism
 refer to [16,17].} In the important special case when the worldlines are hypersurface
 orthogonal the result of this procedure is equivalent to the so called
 $3+1$ or ADM decomposition [18] in which the spacetime is foliated by
 hypersurfaces regarded to be the momentary spaces with the corresponding
 induced metrics as the momentary spatial metrics. The $3+1$ approach despite being very well known
 through the introduction of the total mass and energy in asymptotically flat spacetimes,
 suffers the deficiency of being only fairly applicable to the case of product manifolds.\\
 The threading decomposition leads to the following splitting of the spacetime distance element [19]
$${\rm d}s^{2}={\rm d}T^{2}-{\rm d}L^{2} \eqno(3.1)$$
where the invariants ${\rm d}L$ and ${\rm d}T$ are respectively the \textit{spatial}
and \textit{temporal} length elements of two nearby events as measured by the
threading observers. They are constructed from the normalized tangent vector
$u^{a}={\zeta^a \over |\zeta|}$ to the timelike curves (observer 4-velocities)
in the following way
$${\rm d}L^{2}= -\bar{g}_{ab}{\rm d}x^{a}{\rm d}x^{b}\eqno(3.2)$$
$${\rm d}T= u_{a}{\rm d}x^{a}. \eqno(3.3)$$
Defining $$h = |\zeta|^2$$ $$A_a = -{\zeta_a \over
|\zeta|^2}$$ equations (3.1) and (2.7.b) take the
following forms
$${\rm d}s^{2}=  h(A_a {\rm d}x^a)^2 - h_{ab}{\rm d}x^a
{\rm d}x^b \eqno(3.4.a)$$
$$h_{ab}= -\bar{g}_{ab}= -g_{ab}+h A_{a}A_{b}. \eqno(3.4.b)$$
Given an $\Re-$manifold, a coordinate system is said to be a \textit{preferred} one
if the partial derivative operator with respect to one of the coordinates coincides locally   with the $\zeta$ vector field of the action while the other coordinates label the orbits,
 so that\footnote{Hereafter equations written in this
 preferred coordinate system are denoted by the sign $\doteq$.}
$$\zeta^a \doteq (1,0,0,0)\;\;\; ; \;\;\;
A_a \doteq - {g_{0a}\over g_{00}}\;\;\; ; \;\;\;h=e^{2\nu}\doteq
g_{00}\eqno(3.5)$$ and the above spatial and spacetime distance
elements take the following forms
$${\rm d}L^2 \doteq \gamma_{\alpha\beta}{\rm d}x^{\alpha}{\rm d}x^{\beta} \eqno(3.6.a)$$
$${\rm d}s^{2} \doteq h({\rm d}x^{0}-A_{\alpha}{\rm d}x^{\alpha})^{2}-{\rm d}L^2 \eqno(3.6.b)$$
where
$$\gamma_{\alpha\beta} = -g_{\alpha\beta} +{g_{0\alpha}
g_{0\beta}\over g_{00}}. \eqno(3.7)$$
Also the equations $(2.8.a)$, $(2.8.b)$, $(2.5)$ and $(2.6)$ transform into [11]
$$ ^{\perp}{\partial}_{\alpha} \doteq  \partial_{\alpha} +
A_{\alpha} \partial_{0} \eqno(3.8.a)$$
$$ ^{\perp}{\partial}_{0} \doteq 0  \eqno(3.8.b)$$
$${}^{\perp}{\rm d}^{\alpha} \doteq {\rm d}x^{\alpha} \eqno(3.8.c)$$
$${}^{\perp}{\rm d}^{0} \doteq A_{\alpha}{\rm d}x^{\alpha} \eqno(3.8.d)$$
$${}^{\perp}{\bigtriangledown}_{\beta}X^{\alpha} \doteq X^{\alpha}_{*\beta}+
\Upsilon^{\alpha}_{\beta\eta}X^{\eta}\;\;;\;\;\forall X \in {}^{\perp}TM $$
$$\bar{Z}^{\alpha}_{\beta\eta\rho} \doteq \Upsilon ^{\alpha}_{\beta\rho*\eta}
-\Upsilon^{\alpha}_{\beta\eta*\rho}+
\Upsilon^{\alpha}_{\sigma\eta}\Upsilon^{\sigma}_{\rho\beta}-
\Upsilon^{\alpha}_{\sigma\rho}\Upsilon^{\sigma}_{\eta\beta} \eqno(3.9.a)$$
where
$$\Upsilon^{\alpha}_{\beta\eta} \doteq \frac{1}{2}\gamma^{\alpha\rho}
(\gamma_{\beta\rho*\eta}+ \gamma_{\rho\eta*\beta}-
\gamma_{\beta\eta*\rho}) \eqno(3.9.b)$$
and $*$ stands for ${}^{\perp}\partial$.\\
In this context the velocity of a particle as measured by the threading
observers is given by [14]
$${\cal V}^{\alpha} \doteq \frac{{\rm d}x^{\alpha}}
{\sqrt{h}\;{}^{||}{\rm d}^{0}}\eqno(3.10.a) $$
or covariantly
$${\cal V}^{a} = \frac{{}^{\perp}{\rm d}^{a}}{{\rm d}T}. \eqno(3.10.b)$$
The \textit{spatial} 4-\textit{force} due to the spacetime curvature which
deviates test particles from geodesics
of the orbit manifold and makes them follow the geodesics of
spacetime is defined by [14]
$${\cal F}^{a}= \frac{{}^{\perp}{\cal D}{\cal P}^{a}}{{\rm d}T} \eqno(3.11.a)$$
where\footnote{We use gravitational units
where c=G=1.}
$${\cal P}^{a} = \frac{m}{\sqrt{1-{\cal V}^{2}}}{\cal V}^{a} \eqno(3.11.b)$$
and
$${\cal V}^{2}=h_{ab}{\cal V}^{a}{\cal V}^{b}. \eqno(3.11.c)$$
As is shown in appendix A, by introducing the {\it
gravitoelectric}\footnote{For reviews on the subject of
Gravitoelectromagnetism see references [2] and [20].} and {\it
gravitomagnetic} fields [12] \footnote{Notice that \textit{in this paper} the signs $[,]$ and $(,)$
denote the commutation and anti-commutation over indices,  in example
for a tensor $T_{ab}$, $T_{[ab]}=T_{ab}-T_{ba}$.}
$${\cal E}_{a} = -\nu_{;a}-\zeta^{b}(\nu_{;b}A_{a}+F_{ba}) \doteq (0,\vec{\cal E})
\eqno(3.12)$$
$$$$
$${\cal B}_{ab} = F_{ab}-A_{[a}F_{b]c}\zeta^{c}
\doteq \sqrt{\gamma}
\pmatrix{
 0 & 0 & 0 & 0 \cr
 0 & 0 & {\cal B}^3 & -{\cal B}^2 \cr
 0 & -{\cal B}^3 & 0 & {\cal B}^1 \cr
 0 & {\cal B}^2 & -{\cal B}^1 & 0 \cr}
 \eqno(3.13)$$
where
$$F_{ab} = A_{[\;b\;;\;a\;]} \eqno(3.14)$$
and
$$\vec{{\cal E}} = -(\nabla \nu + \vec{A}_{,0} + \nu_{,0} \vec{A}) \eqno(3.15.a)$$
$$\vec{{\cal B}} = \nabla \times \vec{A} + \vec{A} \times \vec{A}_{,0}\eqno(3.15.b)$$
$\bar{{\cal F}}_{a} = h_{ab}{\cal F}^{b}$ is found to be
$$\bar{{\cal F}}_{a} + f_{a} = \frac{m}{\sqrt{1-{\cal V}^{2}}}(
{\cal E}_{a} + \sqrt{h}\;{\cal B}_{ab}\;{\cal V}^{b})\eqno(3.16.a)$$
where
$$f_{a} = \frac{1}{\sqrt{h}}\pounds_{\zeta}\bar{{\cal P}}_{a} = {2}\frac{m}{\sqrt{1-{\cal
 V}^{2}}} D_{ab}{\cal V}^{b} +
 m_{||}(- \frac{1}{\sqrt{h}}\zeta^{b}\nu_{;b} + \zeta^{c}F_{cb}{\cal V}^{b}+
D_{bc}{\cal V}^{b}{\cal V}^{c}){\cal V}_{a}  \eqno(3.16.b)$$
$$m_{||}=\frac{m}{(1-{\cal V}^{2})^{\frac{3}{2}}}$$
$$\bar{{\cal P}}_{a} = h_{ab}{\cal P}^{b}$$
and [12]
$${\cal D}_{ab}=\frac{1}{2\sqrt{h}}\pounds_{\zeta}h_{ab}. \eqno(3.17)$$
$f_{a}$ can be regarded as a \textit{friction like
force} which appears in the non-stationary cases and the right hand side of the equation
(3.16.a) can be called the \textit{gravito-Lorentz} force.\\
Using $\gamma_{\alpha\beta}$ as the metric tensor,
 one can write the vacuum Einstein field equations for a general spacetime in the following   \textit{quasi-Maxwell} form [12]
$${}^{\perp}{\nabla} . {\vec{{\cal E}}}\;\doteq\;{1\over 2}h {\vec{{\cal
B}}}^{2}\;+\;{\vec{{\cal E}}}^{2}\;
-\;\hat{{\cal D}}\;-\;\frac{1}{\sqrt{h}}\partial_{0}{\cal D}      \eqno(3.18.a)$$
$${}^{\perp}{\nabla}\times(\sqrt{h}\vec{\cal B})\;\doteq\;2\;({\vec {\cal E}} \times
{\sqrt{h}{\vec {\cal B}}}\;+\;{\vec{\cal S}}) \eqno(3.18.b)$$
$$\bar{Z}_{\alpha\beta}\;\doteq\;{\cal E}_{\alpha}{\cal E}_{\beta}
-\frac{1}{2}{}^{\perp}{\bigtriangledown}_{(\alpha}{\cal E}_{\beta)}
 -\frac{h}{2}{\cal B}_{\alpha\eta}{\cal B}^{\bf{.}\eta}_{\beta} +$$ $$
  \frac{\sqrt{h}}{2}({\cal D}_{\alpha\eta}{\cal B}^{\bf{.}\eta}_{\beta}+
  {\cal B}_{\alpha\eta}{\cal D}^{\eta}_{\beta})
   + 2 {\cal D}_{\alpha\eta}{\cal D}^{\eta}_{\beta} - {\cal D}{\cal D}_{\alpha\beta}
    -\frac{1}{\sqrt{h}}\partial_{0}{\cal D}_{\alpha\beta}
 \eqno(3.18.c)$$
where
$${\cal D} = Tr({\cal D}_{ab})\eqno(3.19.a)$$
$$\hat{{\cal D}} \doteq {\cal D}_{\alpha\beta}{\cal D}^{\alpha\beta} \eqno(3.19.b)$$
$$S^{\alpha} \doteq \; {}^{\perp}{\partial}^{\alpha}{\cal D}\;-\;
{}^{\perp}{\bigtriangledown}_{\beta}{\cal D}^{\alpha\beta}$$
and $\bar{Z}_{\alpha\beta}$ is the 3-D Ricci tensor constructed from $\bar{Z}_{\alpha\beta\eta\rho}$.
For a stationary spacetime the gravitoelectric and gravitomagnetic
fields are  curl-free and divergenceless respectively. Therefore by taking
the preferred coordinate system to be the one adapted to
the congruence of its timelike killing vector field $\xi_t=\partial_t$, the field equations
simplify significantly [2].
\section{Gauged Lie derivative}
Now we investigate the expected reparametrization invariance of the threading decomposition.
Considering two vector fields $\zeta$ and $\grave{\zeta}$, the question is whether
they produce the same induced $1+3$ \textit{physics}, in the sense discussed in section
3. The answer comes in two parts, for $\grave{\zeta}$
to reproduce the same threading orbits and orthogonal splitting structure as $\zeta$ i.e.
$$\bar{\grave{M}} = \bar{M}\;\;;\;\; \grave{P} = P \eqno(4.1.a)$$
we only need to have
$$\grave{\zeta} = \Omega \zeta \eqno (4.2.a)$$ where $\Omega$ is a real-valued function on $M$.
However if we further require that $\zeta$ and $\grave{\zeta}$ satisfy
$${\cal E}_{a} = \grave{{\cal E}}_{a}\;\;;\;\;
\sqrt{h}{\cal B}_{ab} = \sqrt{\grave{h}}\grave{{\cal B}}_{ab} \eqno(4.1.b)$$
$$\frac{\zeta^{a}}{|\zeta|} = \frac{\grave{\zeta}^{a}}{|\grave{\zeta|}}\;\;;\;\;
{\cal V}^{a} = \grave{\cal V}^{a} \eqno(4.1.c)$$
along with
$$\frac{h(q_{1})}{h(q_{2})} = \frac{\grave{h}(q_{1})}{\grave{h}(q_{2})}\;\;;\;\;
\forall (q_{1},q_{2}) \in M^{2} \eqno(4.1.d)$$
i.e yield the same spatial force
and the same ratio of the proper observer clock rates at any two arbitrary points,
a more restricted form of the relation $(4.2.a)$ is resulted, that is
$$\grave{\zeta} = \kappa \zeta \;\;;\;\;\forall \kappa \in {\cal R}. \eqno(4.2.b)$$
Now according to $(4.2.b)$ the two parametrizations of the threading orbits corresponding to $\zeta$
and $\grave{\zeta}$ which defined by $\zeta=\partial_{\tau}$ and
$\grave{\zeta} = \partial_{\grave{\tau}}$ respectively, are related to each other by
$$\grave{\tau} = \;\frac{1}{\kappa}\; \tau -\; \phi \eqno(4.3.a)$$
where $\phi$ is any real-valued function on $M$ satisfying the
following condition
$$q_{2} \in \Re.q_{1} \Rightarrow \phi(q_{1}) = \phi(q_{2})\;\;;\;\;
\forall (q_{1},q_{2}) \in M^{2} \eqno(4.3.b)$$
that is the values of $\phi$ are the same along each threading orbit and so
it can be equivalently regarded as a function on $\bar{M}$.\\
As a conclusion the physics of the $1+3$ spacetime splitting is
\textit{quantitatively} invariant under a reparametrization of the threading orbits given
by $(4.3.a-b)$.\\
In this context given an $\Re$-manifold $M$ and two vector fields $X$ and $Y$, we define
the gauged Lie derivative of $Y$ with respect to $X$ by following the
same procedure used to define the standard Lie derivative but incorporating
the previously mentioned reparametrization invariance of the threading decomposition.\\
Considering a point $q \in M$ and a coordinate system $S$, the Lie derivative of $Y$
with respect to $X$ is given by [21]
$$({}_{q}{\pounds_{X}Y})^{a} = {\rm lim}_{\delta\lambda \rightarrow 0}
\frac{{}_{q^{'}}Y^{a}_{|S^{'}} - {}_{q}Y^{a}_{|S}}{\delta\lambda}$$
where $q$ and $q^{'}$ are both on the same $X$-orbit with coordinates
$x^{a}$ and $x^{a} + \delta\lambda X^{a}$ respectively and $S^{'}$ is
another coordinate system which is related to $S$ by
$$x^{'a} = x^{a} - \delta\lambda X^{a}. $$
According to the above definition ${}_{q}\pounds_{X}Y^{a}$ measures the
momentary variation rate
of $Y^{a}$ as seen by a coordinate frame moving along the $X$-orbit at the point $q$. Now
to incorporate the invariance under the reparametrization $(4.3.a-b)$ we try as an alternative
to the usual definition of the Lie derivative the following one
$$({}_{q}^{\phi} {\pounds_{X}Y})^{a} \doteq
{\rm{lim}}_{\delta\lambda \rightarrow 0}
\frac{{}_{q^{'}}Y^{a}_{|\grave{S}} - {}_{q}Y^{a}_{|S}}{\delta\lambda}$$
where $q^{'}$ is the \textit{same point as before}, $S$ is a preferred coordinate system
for the threading orbits and the coordinate system $\grave{S}$ is related to $S$ by
$$\grave{x}^{a} \doteq x^{a} - \delta\lambda X^{a} - \zeta^{a}(\delta\lambda
\phi(x^{\alpha}) + \delta \kappa\;x^{0}).$$
Keeping $\bar{\kappa} = \frac{\delta \kappa}{\delta \lambda}$ constant, the above limit yields
$$^{\phi}\pounds_{X}Y^{a} = \pounds_{X}Y^{a} - \zeta^{a}Y^{b}\phi_{,b}
- \bar{\kappa} \zeta^{a}Y^{0}.$$
However since we request the result of a gauged Lie derivative be a tensor field,
we have to set $\bar{\kappa} = 0$. This choice however does not reduce the generality
of our argument from physical standpoint because $\bar{\kappa}$ introduces
a constant rescaling of the time coordinate. Therefore
$$^{\phi}\pounds_{X}Y^{a} = \pounds_{X}Y^{a} - \zeta^{a}Y^{b}\phi_{;b}
\eqno(4.4.a)$$
defines the \textit{gauged Lie derivative} of a vector field with respect to a doublet,
a vector field and \textit{a gauge}, a real-valued function $\phi \doteq \phi(x^{\alpha})$
 on the $\Re$-manifold under consideration.\\
Demanding that for any real-valued function ${\rm f}$ on $M$
$$^{\phi}\pounds_{X}{\rm f} = \pounds_{X}{\rm f} \eqno(4.4.b)$$
and requesting the gauged Lie derivative to respect the Leibniz rule, for a one-form
$\omega$ we obtain
$$^{\phi}\pounds_{X}\omega_{a} = \pounds_{X}\omega_{a} +
\zeta^{b}\omega_{b}\phi_{;a}.
\eqno(4.4.c)$$
Similarly for tensor fields of type $(1,1)$ and $(0,2)$
$$^{\phi}\pounds_{X}T^{a}_{b} = \pounds_{X}T^{a}_{b} - \zeta^{a}T^{c}_{b}\phi_{;c}
+ \zeta^{c}T^{a}_{c}\phi_{;b} \eqno(4.4.d)$$
$$^{\phi}\pounds_{X}T_{ab} = \pounds_{X}T_{ab} + \zeta^{c}T_{cb}\phi_{;a}
+ \zeta^{c}T_{ac}\phi_{;b}. \eqno(4.4.e)$$ The gauged Lie derivative
of other mixed tensor fields are defined in the same manner. This
new definition will be used in the following sections to introduce the
gauged motion.
\section{Gauged motion in the stationary case}
The first order variation of ${\rm d}s^{2}$ under
$$\delta x^{a} = \delta\lambda\;K^{a} \eqno(5.1)$$
corresponding to the infinitesimal motion
$x^{a} \rightarrow x^{a} + \delta\lambda K^{a}$, generated by a vector field $K^{a}$,
yields [21]
$$\delta({\rm d}s^{2}) = \delta\lambda {\rm d}x^{a}{\rm d}x^{b}\pounds_{K}g_{ab} $$
which vanishes if $K$ is an isometry generator of the spacetime.
To derive the so called gauged motion, in this case we use the same
method as above but apply it to the spatial and temporal line elements
$\;{\rm d}T^2$ and $\;{\rm d}L^2$ given in $(3.1)$.
The formulation should be in such a way that the
reparametrization invariance $(4.3.a-b)$ to be incorporated in the
definition of a {\it physical} symmetry. To achieve this goal
we demand that under the variation $(5.1)$ the following requirements hold.\\
${\bf I}.)\;$$\delta({\rm d}L^2) = 0$ from which we have
$${\pounds_K} h_{ab} = 0. $$
${\bf II}.)\;$$ \delta h = 0 $, so that\\
$${\pounds_K} h = 0 $$
which in turn by $(3.12)$ reduces to
$${\pounds_K} {\cal E}_a = 0.  \eqno(5.2)$$
${\bf III}.)\;$Finally
$${}^{\phi}{\pounds_K}A_{a} = 0 \eqno(5.3) $$
which is demanded by the arguments of the previous section and due to which we have
$${\pounds_K}A_{a}\; =\; \phi_{;a}$$
$${\pounds_K} F_{ab} = 0 \eqno(5.4)$$
where in the stationary case
$${\cal B}^\alpha \doteq \frac{1}{2\sqrt{\gamma}}\epsilon^{\alpha\beta\eta}F_{\beta\eta}.$$
One can obtain the same result by demanding the following requirement
$$ \delta({\rm d}T) = \delta\lambda\;{\rm d}\phi \;\;;\;\;\phi_{,0}\doteq 0. \eqno(5.5)$$
Another hint to the requirements {\bf I}, {\bf II} and {\bf III} comes from the fact that
a time transformation of the form
$$x^0  \rightarrow x^{\prime 0} \doteq x^0 - \phi(x^\alpha)$$ implies
$$A_a \;\; \rightarrow \;\; A_a  + \phi_{,a}$$
while $\gamma_{\alpha\beta}$ and $g_{00}$ remain invariant [1].
We note that in the above discussion the invariance of the gravitomagnetic fields
under the gauged motion was implied \textit{independently} of their appearance
through the gravito-Lorentz force or
the quasi-Maxwell form of Einstein field equations.\\
According to the above arguments all the physical aspects of a stationary spacetime,
 defined through a $(1+3)$ splitting based on its timelike isometry curves,
are invariant under a special kind of spacetime motions called \textit{gauged motions}.
The generator of such a motion is called a gauged Killing vector field and is defined as follows. \\
\textbf{\textit{Definition}}:\\
A \textit{gauged Killing vector field} (GKV) in an stationary spacetime threaded by
 its timelike Killing vector field $\zeta$, is a vector field $K$ satisfying
$${\pounds_K}h=0                       \eqno (5.6.a)$$
$${\pounds_K} h_{ab}=0                 \eqno (5.6.b)$$
$${\pounds_K} A_a = \phi_{;a}          \eqno (5.6.c)$$
$$\zeta^{a} \phi_{;a} = 0.              \eqno (5.6.d)$$
As a consequence of the equations $(3.4.b)$ and $(5.6.a)-(5.6.c)$ we have
$$\pounds_K g_{ab} =  h \phi_{;(a}A_{b)}. $$
Now if the \textit{fact} presented in section {\bf IX} is applied to the set
of equations $(5.6.a)-(5.6.c)$, one arrives at the following equivalent version of
the above definition\footnote{It
should be noted that there are other generalizations of the Killing motion such as the {\it conformal} and {\it homothetic} Killing
motions [22].}.\\
\textbf{\textit{Definition:}}\\
For a stationary spacetime a vector field $K$ generates a
gauged motion (\textit{gauged isometry}) corresponding to a threading family
of timelike isometry comoving observers $\zeta$, if
$${\pounds_K}g_{ab} = h  \phi_{(;a}A_{b)} \eqno(5.7.a)$$
$${\pounds_{\zeta}}K = 0 \eqno(5.7.b)$$
along with
$${\pounds_{\zeta}}\phi = 0. \eqno(5.7.c) $$
It is notable that any Killing vector field
satisfies the equations $(5.7.a)$ and $(5.7.c)$ with $\phi=0$, but to be
a GKV it has to satisfy the condition $(5.7.b)$ as well, i.e. its components
should be time-independent in the preferred coordinate system. Although it
is natural to expect the Killing vector fields
of a stationary spacetime respect this requirement, there are exceptions such as the boost generators of the Minkowski spacetime.
\section{Curvature invariants of stationary spacetimes and the gauged motion}
In this section we investigate the role of gauged motion to manifest the
symmetries of the curvature invariants of a stationary spacetime. First we show that
for such a spacetime in the preferred coordinate system the gauged Killing
vectors are basically the same as the usual Killing vectors apart
from a local shift in their time components. To prove this we
start from equation $(5.7.a)$ in the following form
$$g_{ab,n}K^{n} + g_{nb} K^{n}_{,a} +  g_{an} K^{n}_{,b}
=  h \phi_{;(a}A_{b)}.$$ Now in the preferred coordinate system for $\zeta^{a}$
the above equation takes the form
$$g_{ab,n}K^{n} + g_{0b}( K^{0}_{,a}+ \phi_{,a}) +
g_{a0}( K^{0}_{,b} + \phi_{,b}) + g_{a \alpha}K^{\alpha}_{,b} +
g_{\alpha b}K^{\alpha}_{,a} \doteq 0. \eqno(6.1)$$ Using the fact that
$g_{ab,0} \doteq 0$ and changing the variable to
$$\xi^{a} = K^{a} + \phi \zeta^{a} \doteq K^{a} + \phi \delta ^{a} _{0}  \eqno(6.2)$$
where $\zeta^a \phi $ is the generator of the {\it
shift of the time zero} in the preferred coordinate system, equation $(6.1)$ reduces to
$$g_{ab,n}\xi^{n} + g_{am}\xi^{m}_{,b}+  g_{mb}\xi^{m}_{,a} = 0$$
which is the Killing equation with the $\xi$ as the Killing
vector. The result of the above calculation can be summarized in
the following relation
$${\pounds_K} g_{ab} =  h \phi_{;(a}A_{b)} \Leftrightarrow {\pounds_{\xi}} g_{ab}=0
\eqno(6.3)$$ where
the relation between $\xi^a$ and $K^a$ is given by $(6.2)$. In other
words here again we basically see the interplay between the gauge
freedom in choosing the gravitomagnetic potential and shifting the time zero as mentioned in  the previous section .\\
The relation (6.2) brings the following consequences:\\
$\bf{I}.)\;$ The gauge isometry group of a
stationary spacetime forms a Lie algebra by its definition, then
due to the equations $(5.7.b)$, $(5.7.c)$ and $(6.2)$ for any two GKVs $K$ and $K^{'}$
$$[K,K^{'}] = [\xi,\xi^{'}]$$
so that the Lie algebra of the gauged isometry group of a
stationary spacetime is the same as the Lie algebra of the
corresponding isometry group. We will meet again this fact while
discussing the symmetries of NUT space.\\
$\bf{II}.)\;$\textbf{\textit{Fact}}:\\ For a stationary spacetime
all the curvature invariants of any order are invariant under the
gauged motion.\\ Using the equation $(6.2)$ and $(6.3)$ and noting that
for any invariant $I^{(n)}$ of order $n$ we have
$${\pounds _{\xi}}I^{(n)} = {\pounds _{K + \phi \zeta}} I^{(n)} = 0$$
or
$${\pounds _{K}}I^{(n)} = - {\pounds _{\phi \zeta}}I^{(n)}$$
and so
$${\pounds _{K}}I^{(n)} = -I^{(n)}_{,a} \zeta^a \phi
\doteq -I^{(n)}_{,0} \phi = 0$$ where in the last step we used the fact
that the spacetime under consideration is stationary. So in the case
of stationary spactime
symmetries of the curvature invariants are described through the gauged motion.
\section{NUT-type spaces and their symmetries}
The three known NUT-type spacetimes are\\
a) NUT space\\
b) Cylindrical NUT \\
c) Planar NUT \\
These spacetimes have the common feature of being stationary solutions of
the vacuum Einstein equations and respectively incorporating spherical, cylindrical and
planar gravitoelectromagnetic fields and the spatial metric in their construction. Therefore as a consequence
of the fact presented in the last section, not only their gravitoelectromagnetic fields
but also all their curvature invariants follow the physical
symmetries demanded by the corresponding gauged motions. In the following
we consider the gauged killing vectors of NUT space as main
prototype exhibiting the main features discussed in the previous
sections. Indeed it was some of the peculiar features of NUT space
that motivated this study in the first place. There are two
different interpretations of NUT space due to Misner [4] and
Bonnor [23] which differ significantly in their physical and
geometrical descriptions of the metric. Misner's interpretation
seems to be the dominant one specially after
the rederivation of NUT space presented in [2] where the ideas
of gravtioelectromagnetism play the essential
role [24]. There, a redrivation of NUT solution is achieved by looking for a
stationary spacetime whose spatial metric and the
gravitoelectromagnetic fields respect spherical symmetry.\\ NUT metric has the following four linearly independent
Killing vectors
$$\xi_t = \partial_t$$
$$\xi_1 = \sin\phi \partial_{\theta} + \cos\phi[\cot\theta \partial_\phi +
2l \csc\theta\partial_t]$$
$$\xi_2 = \cos\phi\partial_\theta - \sin\phi[\cot\theta \partial_\phi +
2l \csc\theta\partial_t]$$
$$\xi_3 = \partial_\phi.$$
Now from the above Killing vectors one could obtain two strange
features of NUT space which in a sense describe the same idea.
First of all one could easily see that the above Killing vectors
satisfy the same commutation relations as the Killing vectors of the Schwarzschild
space, namely
$$[ \xi_\alpha ,\xi_\beta] = -\epsilon _{\alpha\beta\gamma}
\xi_\gamma $$
$$[ \xi_\alpha ,\xi_t] = 0$$
where $\alpha, \beta, \gamma =1,2,3$. In other words the isometry group
is $SO(3)\times U(1)$, where $\xi_t$ generates the group $U(1)$ of time
translations. As none of the Killing vectors contain a term in $\partial_r$,
the orbits of the Killing vectors lie on $r=constant$ timelike
hypersurfaces. Misner has shown that these hypersurfaces are topologically
3-spheres ($S^3$)
, in contrast to the $S^2 \times {\cal R}$ topology of the $r=constant$ hypersurfaces
in the
Schwarzschild metric. One of the main motivation of our work was the above
peculiar symmetry
behaviour of NUT space, the fact that by its mathematical appearance  the
NUT space is an axially symmetric spacetime, but the Lie algebra of its
Killing vectors suggests that it is
intrinsically an spherically symmetric spacetime. In fact it is seen that the NUT-type spaces
 are physically spherical, cylindrical and planar
respectively, specially the curvature invariants follow not the
apparent mathematical symmetry of the line element ${\rm d}s^2$ of
the spacetime under consideration, but its physical symmetry
defined through the gauged Killing motion.\\As a conclusion we mention that all
the above arguments
are encoded in the fact that the gauged isometry group of NUT, cylindrical NUT
and planar NUT are $\Re \times SO(3)\;,\;\Re \times U(1)\times \Re$ and
$\Re \times E_{2}$ respectively, where the first $\Re$ is the group of time translation.
 That is it can be checked easily that the generators of each group satisfy the equations
  (5.7.a-c) for the corresponding spacetime.
\section{Some General properties}
Regarding
$$K_{a;b} + K_{b;a} = h \phi_{;(a}A_{b)} \eqno(8.1)$$
it can be seen immediately that unlike the usual Killing vector,
the gauged Killing vector has in general a non-vanishing
(non-constant) {\it expansion} i.e.
$$\Theta = K^a{}_{;a} = \phi_{;a}A^{a}. \eqno(8.2) $$
We will find below that this main difference shows up in different
contexts. Using the relation
$$T_{a;bn} - T_{a;nb} = R^{m}{}_{abn} T_m \eqno(8.3.a)$$
valid for any vector and the symmetries of the Riemann curvature
tensor one obtains the following identity
$$(T_{a;b} - T_{b;a})_{;n} + (T_{n;a} - T_{a;n})_{;b}
+ (T_{b;n} - T_{n;b})_{;a}=0. \eqno(8.3.b)$$ Using (8.1) again, the above
relation for gauged Killing vectors reads
$$(2 K _{a;b} - h \phi_{;(a}A_{b)})_{;n} + (2 K _{n;a} -
h \phi_{;(n}A_{a)})_{;b} + (2 K _{b;n} - h \phi_{;(b}A_{n)})_{;a}
=0. \eqno(8.4)$$ Using equation $(8.3.a)$ and rearranging terms in the
above equation we obtain
$$K_{n;ba} = R_{mabn} K^m + \left[ (h \phi_{;(b}A_{n)})_{;a}
+(h \phi_{;(n}A_{a)})_{;b}- (h
\phi_{;(a}A_{b)})_{;n}\right]\eqno(8.5)$$ which is the neccessary
condition to be satisfied by any gauged Killing vector.
Using the above equation or equation $(8.3.a)$ we find
$$K^a{}_{;ba}=R_{mb} K^m + (K^a{} _{;a})_{;b}=R_{mb} K^m +
(h \phi_{;a}A^{a})_{;b}. \eqno(8.6)$$ We notice that the above
equation compared with the similar equation satisfied by Killing
vectors and {\it homothetic Killing vectors} [25] has an extra term
in the right hand side which is nothing but the gradient of the
{\it expansion} of the gauged Killing vector $K^{a}$. In analogy
with the definition of Killing bivector (KBV) [25] one can define
the {\it Gauged Killing Bivector } (GKBV) by rewriting equation
$(8.1)$ in the following form
$$Q_{ab} = 2\;K_{a;b} - h \phi_{;(a}A_{b)} \eqno(8.7)$$
where
$$Q_{ab} = K_{[a;b]}\eqno(8.8)$$
could be interpreted as the {\it test electromagnetic field} of
any gauged Killing vector field. Using the above definition along
with equation $(8.1)$ one can show that $Q_{ab}$ satisfies
$$Q^{mn}{}{}_{;n} = R^{mn} K_n + \Theta^{;m} -
[h\phi^{;(n}A^{m)}]_{;n}={4\pi} J^m\eqno(8.9)$$ where $J^{m}$ is
the current corresponding to the above defined {\it test
electromagnetic field}. There are two extra contributions to
 this current compared with the current corresponding to
the homothetic Killing bivector [26]. The first term, as expected,
is the reappearance of the gradient of the expansion of the gauged
Killing vector\footnote{Note that the
homothetic Killing vectors, defined by the equation
$\pounds_H g_{ab} = n g_{ab}$,
do have a non zero expansion but it is a constant and so its gradient does
not contribute in $J^{m}$.} and the second one is basically the
divergence of the symmetric part of $K_{n;m}$ which contributes in
the definition of the shear velocity corresponding to the gauged
Killing vector.
\section{Gauged motion in general spacetimes}
Up to now we have defined the gauged motion for a stationary spacetime
assuming that the threading orbits are the orbits of its timelike Killing vector field.
In this section we consider the general case by relaxing this constraint and let the
 spacetime be either non-stationary or stationary but threaded with an arbitrary
  congruence of timelike orbits. Such an
extension becomes physically important in some cases for example when there is a
 non-Killing threading family of timelike orbits suggested by the nature of
  the spacetime itself such as the galaxy worldlines
in cosmological solutions or dust worldlines in the corresponding spacetimes.
 We will find that the new general definition agrees
   with the one given in section {\bf V} when the restrictions hold. More precisely
   we are looking for a definition of spacetime symmetry as realized by an
   arbitrary family of threading observers in a general spacetime.
    A definition which ensures the invariance of the quantities measured by such a family
 or geometrically speaking guarantees the invariance of the structure of the
 corresponding parametric orbit manifold under the motion generated by a
 vector field. Hence considering the spatial force, the distinct role played by $h$ and $h_{ab}$
 along with the relevant reparametrization invariance encoded in the definition of the gauged
 Lie derivative, we demand our definition to satisfy the following criteria.\\
Given a spacetime and a threading timelike vector field $\zeta$, the spacetime
respects the gauged motion generated by a vector field $K^{a}$ if there exists a
real-valued function on $M$, $\phi$, such that
$$\phi(x^{a}) \doteq \phi(x^{\alpha})\;\;;\;\;$$
$$^{\phi}{\pounds_K}h\; =\;{}^{\phi}{\pounds_K}h_{ab}\;=\;{}^{\phi}{\pounds_K}{\cal E}_{a}
\;=\;{}^{\phi}{\pounds_K}\sqrt{h}{\cal B}_{ab} = 0$$
$${}^{\phi}{\pounds_K}\zeta^{b}F_{ba}\;=\;{}^{\phi}{\pounds_K}\zeta^{b}h_{;b}\;=
\;{}^{\phi}{\pounds_K}{\cal D}_{ab} = 0$$
along with the geometrical condition
$$^{\phi}{\pounds_K}P^{a}_{b}=0. $$
The last condition ensures that through gauged motion $K^{a}$ maps the bundles ${}^{\perp}TM$
and ${}^{\perp}T^{*}M$ to themselves , that is
$$P^{a}_{b}Y^{b}=0 \Rightarrow P^{a}_{b}\;{}^{\phi}{\pounds_K}Y^{b} = 0.$$
Also note that the gravitoelectromagnetic fields not only appear in the
spatial force or
the quasi-Maxwell equations but also have geometrical interpretations in the
splitting structure, see equations (A.1) and (A.2).\\
Now definitions $(4.4.a)-(4.4.e)$, equation $(2.7.a)$ and the following relations
$$\zeta^{a}{h}_{ab}=\zeta^{a}{\cal E}_{a}=\zeta^{a}{\cal B}_{ab}=
\zeta^{a}\zeta^{b}F_{ab}=\zeta^{a}{\cal D}_{ab}=0$$
imply
$${\pounds_K}h = 0                 \eqno(9.1.a)$$
$${\pounds_K}h_{ab} = 0            \eqno(9.1.b)$$
$${\pounds_K}{\cal E}_{a} = 0      \eqno(9.1.c)$$
$${\pounds_K}{\cal B}_{ab} = 0     \eqno(9.1.d)$$
$${\pounds_K}(\zeta^{b}F_{ba}) = 0 \eqno(9.1.e)$$
$${\pounds_K}(\zeta^{b}h_{;b}) = 0 \eqno(9.1.f)$$
$${\pounds_{\zeta}}{\cal D}_{ab} = 0       \eqno(9.1.g)$$
$${\pounds_K}P^{a}_{b} = \zeta^{a}\phi_{;b}. \eqno(9.1.h)$$
Not surprisingly we are interested in a more compact definition of the gauged motion.
The procedure to compactify the above definition is given through the following steps.\\
$\bf{I}.)\;$Defining
$$\pounds_{K}A_{a} = X_{a} \eqno(9.2)$$
equations $(9.1.a), (9.1.b)$ and $(3.4.b)$ yield
$$\pounds_{K}g_{ab} = h A_{(a}X_{b)}. \eqno(9.3.a)$$
However one should notice that the equation $(9.3.a)$ by itself is not equivalent
to the set of equations $\{(9.1.a), (9.1.b), (9.2)\}$. The following fact determines the
extra condition required to have such an equivalence.\\
\textbf{\textit{Fact}}:\\
The following two sets of equations are equivalent
$$A= \{(9.1.a),(9.1.b),(9.2)\}$$ $$B=\{(9.3.a), (9.3.b)\}$$
where
$$\pounds_{K}\zeta^{a} = X_{b}\zeta^{b}\zeta^{a}. \eqno(9.3.b)$$ Proof:\\
The first side:$\;A \Rightarrow B)$\\
$$(9.1.b) \Rightarrow {\pounds_K}h_{a0}=0 \Rightarrow h_{ab}K^{b}_{,0}\doteq 0
\Rightarrow P^{c}_{b}K^{b}_{,0}\doteq 0 \Rightarrow P^{c}_{b} {\pounds_{\zeta}}K^{b}=0
\Rightarrow \pounds_{\zeta}K \parallel \zeta$$
$$\Rightarrow \pounds_{\zeta}K^{a}=\lambda \zeta^{a}$$
But
$$(9.2) \Rightarrow X_{0} \doteq A_{a}K^{a}_{,0} \doteq A_{a} \lambda \zeta^{a} \doteq -\lambda $$
The second side:$\;B \Rightarrow A)$
$${\pounds_K}h = {\pounds_K}(\zeta^{a}\zeta^{b}g_{ab}) \doteq
{\pounds_K}g_{00}+ 2g_{0b} {\pounds_K}\zeta^{b} \doteq -2g_{00}X_{0}+ 2g_{00}X_{0} \doteq 0$$
$$$$
$${\pounds_K}A_{a} = {\pounds_K}(\frac{-g_{ab}\zeta^{b}}{h}) \doteq
-\frac{1}{h}({\pounds_K}g_{a0}+g_{ab}{\pounds_K}\zeta^{b})\doteq
X_{a}-\frac{1}{h}(hA_{a}X_{0}-g_{ab}K^{b}_{,0}) $$
$$\doteq X_{a}-\frac{1}{h}(hA_{a}A_{b}K^{b}_{,0}-g_{ab}K^{b,0}) \doteq
X_{a}-\frac{1}{h}h_{ab}K^{b}_{,0} \doteq X_{a}-\frac{1}{h}g_{cd}P^{c}_{a}P^{d}_{b}K^{b}_{,0}$$
$$\doteq X_{a}+\frac{1}{h}g_{cd}P^{c}_{a}P^{d}_{b}{\pounds_K}\zeta^{b} \doteq
X_{a}+\frac{1}{h}\lambda g_{cd}P^{c}_{a}(P^{d}_{b}\zeta^{b}) = X_{a}$$$$$$
$${\pounds_K}h_{ab} = {\pounds_K}(hA_{a}A_{b}-g_{ab}) = hA_{(a}X_{b)}-{\pounds_K}g_{ab}=0.$$
The condition (9.3.b) carries an important geometrical interpretation which could have
been considered as an independent motivation to demand it in the definition of the
gauged motion from the beginning. It originates from the fact that $K^{a}$ as a smooth vector field locally determines its integral curves and corresponding to them defines a one parameter family of diffeomorphisms of the manifold by translating each point along its integral curve passing through that point.
In this respect considering an infinitesimal translation $\epsilon K^{a}$ and a
preferred chart $(O,\varphi)$ the map $\epsilon_{K}$ is defined by
$$\epsilon_{K}: M \rightarrow M $$
$$\epsilon_{K}(q) \doteq \epsilon_{K}[\varphi^{-1}(x^{a})] \doteq \varphi^{-1}(x^{a} + \epsilon K^{a})\;\;;\;\;
\forall q \in O$$
Now since due to $(9.3.b)$
$$K^{\alpha}_{,0} \doteq 0$$
the map $\epsilon_{K}$ respects the equivalence relation by which $\zeta-$orbits define the orbit manifold $\bar{M}$, that is
$$q \sim q^{'} \Leftrightarrow {\epsilon}_K(q)\;\sim\;\epsilon_{K}(q^{'})\;\;;
\;\;\forall (q,q^{'})\in M^{2}. $$
This is also valid for a finite translation generated by $K^{a}$ and its
corresponding map $\epsilon_{K}$ which  maps $\zeta$-orbits (and hence threading observers)
 to themselves, or simply
$${\epsilon}_K( \Re.q )\; = \;\Re.{\epsilon}_K (q)\;\;;\;\;\forall q\in M.$$
We also note that the condition (9.3.b) is satisfied with vanishing $\lambda$ by all the groups
 of space symmetries (like the gauge group $SO(3)$ in NUT space) and so it is not
a very restrictive condition.\\
$\bf{II}.)\;$Equations $(9.1.h), (9.3.b)$ and $(2.7.a)$ yield
$$X_{a} = {\pounds_K}A_{a} = \phi_{;a} \eqno(9.3.c)$$
or equivalently
$$^{\phi}{\pounds_K}A_{a} = 0.$$
Now the conditions $(9.3.a)$ and $(9.3.b)$ take the forms
$$^{\phi}{\pounds_K}g_{ab} = 0$$
$${\pounds_K}\zeta^{a} = 0.$$
$\bf{III}.)\;$The conditions $(9.1.e)$ and $(9.1.g)$ are automatically satisfied due to the
equations $(9.3.c), (9.3.b)$ and $(9.1.b)$.\\
${\bf IV}.)\;$Recalling the definition $(3.12)$, the conditions $(9.1.a), (9.1.c), (9.1.e)$
 and $(9.1.f)$ imply
$$\nu_{,0}\phi_{,a} \doteq 0$$
so that to let a gauged isometry be \textit{non-trivial}, that is different from an isometry,
we need
$${\pounds_{\zeta}} |\zeta| = 0. \eqno (9.3.d)$$
The above condition which can be replaced with the condition (9.1.c), is interesting in the
sense that it is independent of the vector field $K$. It determines the families of
threading observers which are allowed to be considered for having a gauged isometry.
Physically, this condition in the preferred coordinate system of the threading observers
 $(g_{00,0} \doteq 0)$ states that the rate of each observer's clock
measured by him has to be a constant. Hence the \textit{proper clock rate} of any observer
in a threading family which respects a nontrivial gauged motion is the same everywhere on the
observer worldline.\\
$\bf{V}.)\;$Recalling the definition $(3.13)$ the conditions $(9.1.d), (9.1.e)$ and (9.3.c) imply
$$\nabla \phi \times \vec{A}_{,0} \doteq 0 \eqno(9.3.e)$$
that is
$$\phi_{;a} = \Omega \zeta^{b}F_{ba} \eqno(9.3.f)$$
where $\Omega$ is a real-valued function on $M$. This condition generally
 puts restriction on the spacetimes
capable of admitting gauged motion as it implies that the vector field
 $g^{ab}\zeta^{c}F_{cb}$ is orthogonal to the hypersurfaces of constant $\phi$,
an example of which is when
$${\pounds_{\zeta}}A_{a} \doteq A_{a,0}= 0.$$
The above arguments can be summarized as follows\\
\textbf{\textit{Definition}}:\\
Given a spacetime with a timelike vector field $\zeta$, a vector field $K$ is the generator of a \textit{gauged motion (gauged isometry)} for the corresponding observers if
$${\pounds_K}g_{ab} = \phi_{(;a}\zeta_{b)}     \eqno(9.4.a)$$
$${\pounds_K}\zeta^{a} = 0                     \eqno(9.4.b)$$
$${\pounds_{\zeta}}\zeta_{a}=\Omega \phi_{;a}    \eqno(9.4.c)$$
$$\zeta^{a}\phi_{;a} = 0.                         \eqno(9.4.d)$$\\
It is notable that if one relaxes the condition $(9.4.d)$ and defines the gauged
 motion with a more general gauge $\phi \doteq \phi(x^{a})$, then
the corresponding gauged Lie derivative satisfies the same equations as $(4.4.a-d)$.
Further to that the requirements $(9.1.a-g)$ are retained
but the equation $(9.1.h)$ is replaced with
$$\pounds_{K}P^{a}_{b} = \zeta^{a}(\phi_{;b}+A_{b}\phi_{;c}\zeta^{c}) \eqno(9.1.h^{'})$$
and one finds
$\phi_{,0} \doteq 0$ or $\zeta$ is a timelike Killing vector field. Therefore
the gauged motion is defined through the equations (9.4.a-c) and
the  following equation
$$\phi_{;c}\zeta^{;c}\;{\pounds_{\zeta}}g_{ab} = 0. \eqno(9.4.d^{'})$$\\
Now we ask if there is an \textit{almost stationary} sapcetime in the sense that
despite being non-stationary, it is observationally stationary to the corresponding observers
of a timelike vector field. The answer is negative. To Show this fact starting from
$$\zeta^{a}=K^{a}$$
the equations $(9.1.a)$, $(9.1.b)$ and $(9.3.c)$ yield
$$g_{00,0} \doteq 0$$
$$\gamma_{\alpha\beta,0} \doteq 0$$
$$A_{a,0} \doteq \phi_{,a}\;\;\Rightarrow\;\;
A_{a} \doteq t \phi_{,a}(x^{\alpha}) + {\rm f}_{a}(x^{\alpha})$$
so that by the following redefinition of the timelike coordinate
$$t \rightarrow t^{'} \doteq te^{-\phi}$$
the metric takes the form of a stationary spacetime line element
$${\rm d}s^{2} \doteq e^{2(\nu+\phi)}({\rm d}t^{'}-e^{-\phi}{\rm f}_{\alpha}{\rm d}x^{\alpha})
-\gamma_{\alpha \beta}{\rm d}x^{\alpha}{\rm d}x^{\beta}.$$
The other question is whether in the general time-dependent case the gauged
symmetries coincide with the symmetries of the curvature invariants. The answer
is again negative. One can show that unlike the stationary case, in general
the curvature invariants do not
respect the gauged motion. As a proof, we consider the case in which the generator
of gauged motion, the vector field $K^{a}$, is spacelike. Then as $(9.4.b)$
 holds, the vector fields $\zeta$ and $K$ can define simultaneously a timelike and a
 spacelike coordinate and so a coordinate system can be a preferred one for  both of them.
 Sitting in such a frame
$$\zeta^{a}\doteq (1,0,0,0)$$
$$K^{a}    \doteq (0,1,0,0)$$
and equations $(9.4.a-d)$ are satisfied by a line element such as
$${\rm d}s^{2} \doteq e^{2y}({\rm d}t-(x-t){\rm d}y)^{2}
-e^{t}({\rm d}x^{2}+{\rm d}y^{2}+{\rm d}z^{2}).$$ Now if all the curvature
invariants of this metric satisfy
${\pounds _K}I^{(n)}=0$, each one of them must be $x^{1}$-independent in this
 coordinate system, but this is
not the case as the Ricci scalar shows.\footnote{Here we have used Maple tensor package.
Note also that being a counter-example, the physical significance of this metric is not
 important here.}
That is, despite the fact that the parametric orbit manifold
is gauged invariant, the curvature invariants of spacetime are not generally so.
Despite the above fact, in the next section we show that for Kaluza-Klein theories one can
 generalize the idea of gauged motion to make it capable of describing symmetries of
 both the physical quantities and the curvature invariants of the corresponding 4-D spacetime.
\section{Kaluza-Klein theories and the extended gauged motion}
One way to construct the 5-D Kaluza-Klein theories [27] is to start
from a triplet structure $(^{5}M,g,\zeta)$, a 5-D
pseudo-Riemannian manifold together with a non-null vector field
$\zeta$ whose orbits make a congruence of
smooth curves on $M$. Usually $\zeta$ is taken to be spacelike otherwise
$(^{5}M,g)$ will admit two timelike dimensions. Compact or non-compact versions of the theory are built by taking $\zeta$-orbits to be all $S^{1}$ or ${\cal R}$ where in the
latter case the \textit{cylinder} condition does not hold, i.e the metric components can
depend explicitly on the parameter of the $\zeta$-orbits, denoted by $\iota$. This is so because in Space-Time-Matter (S.T.M) theories [28], from a pure geometry in 5-D, one
 obtains a variety of induced matter fields in 4-D besides the
electromagnetic field. In both versions to identify a 4-D manifold
as the ordinary spacetime, one can let each $\zeta$-orbit collapse
to a singular point and take the corresponding parametric orbit
manifold to be the ordinary spacetime.\footnote{ If the manifold is diffeomorphic to $\Re \times \Sigma$ or $S^{1} \times \Sigma$ with $\Sigma$ admitting
 a Lorenzian induced metric, one can take $\Sigma$ as the 4-D spacetime.} We denote the line
element by ${\rm d}S^{2} = g_{\hat{a} \hat {b}} {\rm d}x^{\hat {a}}{\rm d}x^{\hat {b}}$
where
$x^{\hat {a}} \in \{x^{0},x^{1},x^{2},x^{3},\iota\}$. The metric tensor is usually taken to be
 a solution of the vacuum Einstein field equations in five dimensions with or without the
 cosmological constant.
Considering the orthogonal splitting structure induced by $\zeta^{\hat{a}}$ and defining
$$|\zeta|^2 = \epsilon e^{2\Phi}\;\;;\;\;\epsilon^{2}=1 $$
$$A_{\hat {a}} = -\frac{\zeta_{\hat {a}}}{|\zeta|^{2}} = -\epsilon e^{-2\Phi}\zeta_{\hat {a}}$$
the corresponding projection tensor is
$$P^{\hat{a}}_{\hat {b}} = -\delta^{\hat{a}}_{\hat {b}} + \zeta^{\hat {a}}A_{\hat {b}}$$
and therefore in a preferred coordinate system for $\zeta^{\hat {a}}$ we have
$$ ^{\perp}{\partial}_{a}  \doteq \partial_{a}+ A_{a}\partial_{\iota}$$
$${}^{\perp}{\rm d}^{a} \doteq {\rm d}x^{a}$$
$${}^{\perp}{\bigtriangledown}_{a}X^{c}\; \doteq \;{}^{\perp}{\partial}_{a}X^{c} + \Upsilon^{c}_{ab}X^{b}\;\;;\;\;\forall X \in {}^{\perp}TM$$
$$\Upsilon^{c}_{ab}\doteq \frac{1}{2} \gamma^{cd} ({}^{\perp}{\partial}_{b}\gamma_{ad}
+{}^{\perp}{\partial}_{a}\gamma_{db}-{}^{\perp}{\partial}_{d}\gamma_{ab})$$
and
$${\rm d}S^{2} \doteq \epsilon e^{2\Phi}({\rm d}\iota-A_{a}{\rm d}x^{a})^{2}-{\rm d}s^{2}$$
where
$${\rm d}s^{2} \doteq \gamma_{ab}{\rm d}x^{a}{\rm d}x^{b}$$
is the line element of the corresponding parametric orbit manifold. Therefore the metric tensor of the $4$-D spacetime
 is $\gamma_{ab}$ instead of $g_{ab}$, a fact which was pointed out by Einstein and Bergmann
 by requesting the $4$-D metric tensor to be invariant under a shift of $\iota$'s zero point [29].\\
In this respect the corresponding 4-\textit{velocity} and 4-\textit{force} of particles are defined respectively [30]
$$u^{a} = \frac{{}^{\perp}{\rm d}^{a}}{{\rm d}s} \doteq  \frac{{\rm d}x^{a}}{{\rm d}s}
\eqno(10.1)$$
$${\cal F}^{a} = \frac{{}^{\perp}{\cal D}u^{a}}{{\rm d}s} \eqno(10.2)$$
where $m$ is the rest mass of the particle and ${\rm d}s=\sqrt{|{\rm d}s^{2}|}$. Derivation of the above force is similar to the case given in the appendix and a straightforward        calculation yields
$${\cal F}^{a}+ \Xi m \partial_{\iota}u^{a} \doteq m( \frac{{\rm d}u^{a}}{{\rm d}s}+ \Upsilon^{a}_{bc}u^{b}u^{c}) \eqno(10.3)$$
where
$$\Xi \doteq \frac{{\rm d}\iota}{{\rm d}s} - A_{b}u^{b}.  \eqno(10.4)$$
Assuming that the 5-D spacetime is free of non-gravitational interactions,
a test particle worldline is given by
$$\frac{{\rm d}^{2}x^{\hat{a}}}{{\rm d}s^{2}} + \Gamma^{\hat{a}}_{\hat{b}\hat{c}}
\frac{{\rm d}x^{\hat{b}}}{{\rm d}s}\frac{{\rm d}x^{\hat{c}}}{{\rm d}s}
\doteq\;\frac{S^{''}}{S^{'}}u^{\hat{a}} \eqno(10.5)$$
where
$$S^{'}=\frac{{\rm d}S}{{\rm d}s}\;\;;\;\;S^{''}=\frac{{\rm d}S^{'}}{{\rm d}s}.$$
Equations (10.3) and (10.5) imply
$${\cal F}^{a}\;+\; \Xi\;m \partial_{\iota}u^{a} \doteq m\; [\;term\;1+\;term\;2\;+\;term\;3\;+\;term\;4\;]$$
in which
$$term\;1= -\Xi^{2}\; \Gamma^{a}_{\iota\iota}$$
$$term\;2= -2\;\Xi\; (\Gamma^{a}_{\iota b}+ \Gamma^{a}_{\iota\iota}A_{b})u^{b} $$
$$term\;3= (\Upsilon^{a}_{bc}-2\Gamma^{a}_{\iota b}A_{c}-\Gamma^{a}_{\iota\iota}A_{b}A_{c})
u^{b}u^{c}$$
$$term\;4= \frac{S^{''}}{S^{'}}u^{a}.$$
Rewriting the symbols $\Gamma^{a}_{\hat{b}\hat{c}}$ in terms of the projected variables,
and using the fact
$$\partial_{\iota}u^{a} = {\rm d}x^{a}\partial_{\iota}(ds^{-1}) \doteq
-\frac{1}{2}\gamma_{bc,\iota}
u^{c}u^{b}u^{a}$$ we obtain
$$\bar{{\cal F}}_{a} = \gamma_{ab}{\cal F}^{b} \doteq m\; [\;\Xi^{2} \epsilon e^{2\Phi}\Theta_{a}+ \Xi (\epsilon
e^{2\Phi} \bar{F}_{ab}-\gamma_{ab,\iota})u^{b} +
(\frac{1}{2}\Xi \gamma_{bc,\iota}u^{b}u^{c} + \frac{S^{''}}{S^{'}})u_{a}\;] \eqno(10.6)$$
where
$$\Theta_{a} = -[\Phi_{,a}+A_{a,\iota}+A_{a}\Phi_{,\iota}] $$
or covariantly
$$\Theta_{\hat{a}}=-[\;\Phi_{;\hat{a}}+\zeta^{\hat{c}}(F_{\hat{c}\hat{a}}+
A_{\hat{a}}\Phi_{;\hat{c}})\;] \eqno(10.7)$$
is the \textit{dilaton} field strength and
$$\bar{F}_{ab} = A_{[\;b\;;\;a\;]} + A_{[\;\;a}A_{\;b\;],\iota}$$
or covariantly
$$\bar{F}_{\hat{a}\hat{b}} = F_{\hat{a}\hat{b}} -
A_{[\;\hat{a}} F_{\;\hat{b}\;]\;\hat{c}}\zeta^{\hat{c}} \eqno(10.8)$$
is the \textit{extended Maxwell} tensor field.\\
Therefore due to similar arguments given in the previous section, the
\textit{extended gauged motion} is defined as follows\\
\textbf{\textit{Definition}}:\\
A vector field $K$ is the generator of a gauged motion in a 5-D Kaluza-Klein
theory if
$${\pounds_K} \Phi = 0 \eqno(10.9.a)$$
$${\pounds_K} h_{\hat{a}\hat{b}} = 0 \eqno(10.9.b)$$
$${\pounds_K} \Theta_{\hat{a}} = 0 \eqno(10.9.c)$$
$${\pounds_K} \bar{F}_{\hat{a}\hat{b}} = 0 \eqno(10.9.d)$$
$${\pounds_K} \Xi = 0 \eqno(10.9.e)$$
$${\pounds_K} \frac{S^{''}}{S^{'}} = 0 \eqno(10.9.f)$$
along with
$${\pounds_K}A_{\hat{a}} = \phi_{;\hat{a}}. \eqno(10.9.g)$$
One can again obtain a more compact version of the above definition as before which we
leave it to the interested reader.
\section{Comparison with the case in QFT}
As we know one can find the force-free equation of motion (geodesic equation)
 by the following variational principle
$$\delta \int {\rm d}\lambda \frac{{\rm d}s}{{\rm d}\lambda}= 0$$
in other words $\frac{{\rm d}s}{{\rm d}\lambda}$ or its squared plays the role of the test particle
Lagrangian. Now in some of the solutions of the Einstein field equations 
(such as the NUT-type spaces) we have cases in which the  Lagrangian,
due to the presence of the gravitomagnetic potential, has a smaller
symmetry group than the resulting physical states (described
through the gravitoelectromagnetic fields). This is in contrast to the
case of spontaneous symmetry breaking in QFT where states respect a smaller
symmetry group than the one presented in the Lagrangian . To shed
more light on the comparison consider the explicit example of the non-Abelian 
gauge theory with vector fields in the fundamental representation of $SO(n)$
group given by the following Lagrangian 
$${\cal L} = \frac{1}{2}(\partial_{a}\phi \partial^{a}\phi - V(\phi))$$
where $$\phi =\pmatrix{
 \phi_1  \cr
 \phi_2  \cr
 ... \cr
 \phi_n\cr}$$
 and
 $V(\phi) = \frac{\lambda} {4!} {(\phi^2 -c^2)}^2$ [31]. Minima of the potential occur on $S^{n-1}$
 with the radius $c$ defined by $\phi^2 = c^2$, i.e. they form the set 
 ${\cal M} = \{\phi \; | \; V(\phi) = 0 \}$ . Now for a point $m \in {\cal M}$ the 
 stability group is the subgroup
 $$ H_m = \{ h \in SO(n)\; | \; hm =m \}$$
 under which the minimum value of the potential remains 
 invariant and so it is the residual symmetry group of the broken theory.
 We note the similar reduction to the coset space structure (section {\bf II}) as in 
 the present case ${\cal M} = \frac {SO(n)}{SO(n-1)} = S^{n-1}$ . 
 Therefore the vacuum state does not share the 
 symmetry group $SO(n)$ of the Lagrangian of the theory and breaks it spontaneously.
On the other hand, in the case of general relativity, consider a test particle
moving in the field of a massive object given by the line
element ${\rm d}s^2$. 
As the test particle only feels the
gravitoelectromagnetic fields and the corresponding extra forces, it will
 share their geometrical symmetries and not the one presened by 
 the line element (Lagrangian). Although this may not be a good
comparison as in one side we have a classical theory and on the
other side an intrinsically quantum mechanical theory,
nevertheless one should not ignore the possibility that, in the
case of gravity, this character may be retained in the final
theory of \textit{quantum gravity}.
\section{Conclusions}
A new definition of spacetime symmetry in general relativity which is properly applicable to
spacetimes endowed with a \textit{preferred} timelike vector field is presented in a covariant form.
 Such a preference could be suggested geometrically from symmetry considerations or
  physically from the energy-momentum tensor content of the system.
The definition has the advantage of manifesting some of the hidden physical symmetries of these
 spacetimes.\\
The extended version of the definition in higher dimensional
spacetimes with a preferred non-null vector field, presented here for the case of 5-D
Kaluza-Klein theories, could be expected to find application in the various
methods of quantizing gravity for the following two reasons.\\
Firstly the spacetime splitting procedure used here to define the gauged motion is
not restricted to the case of globally product manifolds and this is important in the
sense that in quantum gravity the spacetime topology is
expected to be a dynamical entity.\\
Secondly, by its definition, gauged motion focuses on the physical aspects of a spacetime,
 something reminiscent of the idea of observables in quantum field theories, moreover
this new definition is more flexible than the standard one by admitting a gauge
 \textit{freedom}.
\section*{acknowledgements}
The authors would like to thank university of Tehran for supporting
this project under the grants provided by the research council.
We would also like to thank Prof. D. Lynden-Bell for his useful
comments on the first draft of the paper and for kindly providing us 
with his rewrite of Zelmanov equations in the quasi-Maxwell form which we used to compare with our results in section {\bf III}. We 
also thank Prof. M. Shahshahani for his valuable comments on section {\bf II} 
and M. Haddadi and M. Yavari for useful discussions.
\section*{APPENDIX: Gravitoelectromagnetic fields, the spatial force
and geometry of splitting}
According to the definitions $(3.11.a)$ and $(3.11.b)$
$${\cal  F}^{\alpha} \doteq \frac {m}{{\sqrt h}\;\;{}^{||}{\rm d}^{0}}{}^{\perp}{\cal D}
 \frac{{\cal V}^{\alpha}}{\sqrt{(1-{\cal V}^{2})}}$$
in which
$${}^{||}{\rm d}^{0} \doteq {\rm d}x^{0}-A_{\beta}{\rm d}x^{\beta}
\doteq \frac {{\rm d}s}{\sqrt{h}\sqrt{(1-{\cal V}^{2})}}.$$
Using the definition $(2.8.c)$
$${}^{\perp}{\cal D}{\cal P}^{\alpha} \doteq {\rm d}x^{\beta}\;(
{}^{\perp}{\partial}_{\beta}{\cal P}^{\alpha}+\Upsilon^{\alpha}_{\beta \eta}
{\cal P}^{\eta})$$
so that
$${\cal F}^{\alpha}+\frac{1}{\sqrt{h}}{\cal P}^{\alpha}_{,0} \doteq \sqrt{(1-{\cal V}
^{2})}\frac{{\rm d}}{{\rm d}s}
\frac{m{\cal V}^{\alpha}}{\sqrt{1-{\cal V}^{2}}} + \Upsilon^{\alpha}_{\beta\eta}
\frac{m{\cal V}^{\beta}{\cal V}^{\eta}}{\sqrt{1-{\cal V}^{2}}}.$$
Now due to the fact that particles follow the spacetime geodesics
 and also by the definition $(3.10.a)$
$$u^{\alpha}\doteq\frac{{\cal V}^{\alpha}}{\sqrt {1-{\cal V}^{2}}}$$
$$u^{0}\doteq\frac{1}{\sqrt{1-{\cal V}^{2}}}(\frac{1}{\sqrt{h}}+
A_{\beta}{\cal V}^{\beta})$$
replacing $\frac{{\rm d}u^{\alpha}}{{\rm d}s}$ by $-\Gamma^{\alpha}_{bc}u^{b}u^{c}$
 yields
$${\cal  F}^{\alpha}+\frac{1}{\sqrt{h}}{\cal P}^{\alpha}_{,0} \doteq
\frac{m}{\sqrt{1-{\cal V}^{2}}}[\;term\; 1\\+\;term\;2\;+\;term\;3\;]$$ where
$$term\;1=-\frac{\Gamma^{\alpha}_{00}}{h}$$
$$term\;2=-\frac{2}{\sqrt{h}}(\Gamma^{\alpha}_{0\beta}+\Gamma^{\alpha}_{00}
A_{\beta}){\cal V}^{\beta}$$
$$term\;3=-(\Gamma^{\alpha}_{\beta\eta}+\Upsilon^{\alpha}_{\beta\eta}
+2\Gamma^{\alpha}_{0\beta}A_{\eta}+\Gamma^{\alpha}_{00}A_{\beta}A_{\eta})
{\cal V}^{\beta}{\cal V}^{\eta}.$$
Now using the relations
$$g_{\beta \eta} \doteq -\gamma_{\beta \eta} + hA_{\beta}A_{\eta}$$
$$g^{\beta \eta} \doteq -\gamma^{\beta \eta}$$
$$g^{0\beta} \doteq -\gamma^{\beta\eta}A_{\eta} \doteq -\bar{A}^{\beta}$$
and rewriting the quantities $\Gamma^{\alpha}_{bc}$ in terms of the
spatial metric, the gravitomagnetic vector potential and their derivatives
one finds that the term 3 vanishes and
$${\cal  F}^{\alpha}+\frac{1}{\sqrt{h}}{\cal P}^{\alpha}_{,0} \doteq$$
$$\frac{m}{\sqrt{1-{\cal V}^{2}}}[-(\frac{h^{\alpha}}{2h}+\gamma^{\alpha\beta}
A_{\beta,0}
+\frac{h_{,0}}{2h}\bar{A}^{\alpha})+({\sqrt{h}}(A^{,\alpha}_{\beta}+\bar{A}^{\alpha}
A_{\beta,0}
-\gamma^{\alpha\eta}A_{\eta,\beta}
-\gamma^{\alpha\eta}A_{\eta,0}A_{\beta})
-\frac{1}{\sqrt{h}}\gamma^{\alpha\eta}\gamma_{\eta\beta,0}){\cal V}^{\beta}].$$
Contracting ${\cal F}^{\alpha}$ with $\gamma_{\mu\alpha}$ one obtains
$$\bar{{\cal F}}_{\mu}+f_{\mu}=\frac{m}{\sqrt{1-{\cal V}^{2}}}[
{\cal E}_{\mu}+({\vec{\cal V}}\times\sqrt{h}{\vec{\cal B}})_{\mu}]$$
where \footnote{
$\epsilon^{\alpha\beta\eta}$ is the 3D-permutation pseudo-tensor times the factor
$\frac {1}{\sqrt{\gamma}}$, in terms of which the vector product and the curl operator
are defined.}
$$ {\cal E}_{\mu}=-(\nu_{,\mu}+A_{\mu,0}+\nu_{,0}A_{\mu})$$
$$ {\cal B}^{\alpha}
\doteq \epsilon^{\alpha\beta\eta}\;{}^{\perp}{\partial}_{\beta}A_{\eta}
\doteq {}^{\perp}\nabla \times {\vec A} \doteq \frac{1}{2}
\epsilon^{\alpha\beta\eta}\;{\cal B}_{\beta\eta}$$
 and
$$f_{\mu} \doteq \frac{1}{\sqrt{h}}{\cal P}_{\mu,0}$$
$$\doteq \frac{m}{\sqrt{1-{\cal V}^{2}}}\frac{1}{\sqrt{h}}
\gamma_{\mu \beta ,0}{\cal V}^{\beta}+
m_{||}(A_{\beta,0}{\cal V}^{\beta}+\frac{1}{2\sqrt{h}}\gamma_{\alpha\beta,0}
{\cal V}^{\alpha}{\cal V}^{\beta}
-\frac{1}{\sqrt{h}}\nu_{,0}){\cal V}_{\mu}$$ is the extra force,
a characteristic of non-stationary spacetimes. The above
quantities can be defined covariantly as $(3.12)$, $(3.13)$, $(3.16.a)$, $(3.16.b)$
and $(3.17)$.\\
It is notable that the gravitoelectromagnetic fields can be defined (both
physically and geometrically) independently
from their appearance in the spatial force. Physically since
 ${\cal E}_{a}$ and $\frac{\sqrt{h}}{2}{\cal B}_{ab}$ are
the acceleration and the rotational velocity tensor fields of
$\zeta$-observers respectively and geometrically through the following equations [12]
$$[{}^{\perp}\partial_{a},{}^{\perp}\partial_{b}] \doteq {\cal B}_{ab}\partial_{0}
\eqno(A.1)$$
$$[{}^{\perp}\partial_{a},\frac{1}{\sqrt{h}}\partial_{0}] \doteq {\cal
E}_{a}\partial_{0}.
\eqno(A.2)$$
Hence a non-zero ${\cal B}_{ab}$ implies that the basis $^{\perp}\partial_{a}$ does not define a coordinate basis, a fact which also reflects itself in the following argument.\\
The orbits of $\zeta$ are hypersurface-orthogonal
if and only if there are two real-valued functions ${\rm f}$ and $\Omega$
on $M$ such that
$$A_{a}=\frac{\zeta_{a}}{|\zeta|^{2}}=\Omega {\rm f}_{,a} \eqno(A.3)$$
where the hypersurfaces satisfy ${\rm f}(x^{a})=$constant.
An equivalent neccessary and sufficient condition to $(A.3)$ is
$$A_{[a} {\bigtriangledown}_{b} A_{c]} = 0. \eqno(A.4)$$
Now regarding the definition $(3.13)$
$$A_{[a} {\bigtriangledown}_{b} A_{c]} = A_{a}{\cal B}_{bc} + A_{b}{\cal B}_{ca} +
 A_{c}{\cal B}_{ab} $$
$$A_{[0} {\bigtriangledown}_{b} A_{c]} \doteq {\cal B}_{bc}  $$
and therefore
$$A_{[a} {\bigtriangledown}_{b} A_{c]} = 0\;\;\Leftrightarrow\;\;{\cal B}_{ab} =0 \eqno(A.5)$$
 i.e a vector field is hypersurface orthogonal if and only if the
corresponding gravitomagnetic field vanishes.

\end{document}